# Signatures of Hit and Run Collisions

**Erik Asphaug**
Arizona State University
Chapter for *Planetesimals: Early Differentiation and Consequences for Planets*
(L. T. Elkins-Tanton and B. P. Weiss, eds.), Cambridge University Press 2016

## 1. Introduction

Terrestrial planets grew in a series of similar-sized collisions (SSCs; see **Table 1**) that swept up most of the next-largest bodies. Theia was accreted by the Earth to form the Moon according to the theory. Planetesimals likewise may have finished their accretion in a sequence of 'junior giant impacts', scaled down in size and velocity. This chapter considers the complicated physics of pairwise accretion, as planetesimals grow to planetary scales, and considers how the inefficiency of that process influences the origin of planetesimals and the diversity of meteorites and primary asteroids.

Simulations of planetary collisions show that they are imperfect mergers. Accretion inefficiency gets concentrated, as it were, in the unaccreted bits and pieces, giving asteroids and meteorites their distinctive record according to the arguments outlined below. Mass asymmetry moves the collision point off-center, causing angular momentum asymmetry; a massive target $M_1$ therefore serves both as an 'anvil' into which a next-largest-body (NLB) collides, and as a gravitational pivot around which it gets swung. For characteristic accretion velocities the system can be gravitationally bound, or not. The result is a complicated dynamical process whose outcomes are diverse, depending on normalized projectile mass $\gamma=M_2/(M_1+M_2)$, initial velocity $v_{rel}$, mutual escape velocity $v_{esc}$ (which scales linearly with size), impact angle $\theta$ (median value 45°), compositional differentiation, and rotational and thermal state.

The standard model of Moon formation is an effective accretion of a Mars-like planet by the proto-Earth, starting from $v_{rel}\sim0$ (that is, impact velocity $v_{imp}\sim v_{esc}$). But the more general case, in the absence of drag, is a population of planetary bodies that becomes excited by close mutual gravitational encounters so that faster impacts are common. Characteristic relative velocities increase to $\sim v_{esc}$, and about half of SSCs end up with too much angular momentum and too much energy to result in effective merger. Mantle-stripped cores, stranded clumps, and dispersed sheets become the norm, broadly classified as hit and run collisions (HRCs) when most of $M_2$ escapes downrange.

As described below, a typical HRC results in an escaping core fragment $M_2'<M_2$ accompanied by clumps or arms of its disrupted mantle and crustal material. Other HRCs are super-catastrophic, transforming $M_2$ into an escaping spiral arm or plume. This aspect of pairwise accretion, the mantle stripping and catastrophic disruption of $M_2$ during HRC collisions, can explain how massive planetesimals were destroyed during the formation of planets. Furthermore, the HRC hypothesis can explain how these disruption byproducts disappeared, the 'missing mantle' paradox (Burbine et al. 1996).

Consider an idealized HRC that produces one stripped core ($M_2'$), a dozen mantle clumps, and thousands of bits. The orbits of all these bodies intersect, and their cross section is high, so their ongoing accretion is likely. But sweep-up is strongly biased to favor the most massive object, $M_1$. Consider a size-independent sweep-up happening at random to 90% of remnants. Nine times out of ten, $M_2'$ is accreted by $M_1$ and disappears. In this case there is one 'orphaned' mantle clump (on average) and hundreds of bits, but no parent asteroid. Lots of orphans are predicted by this hypothesis. One time in ten, $M_2'$ is not swept up; it becomes part of an increasingly diverse population of stripped original bodies. Although not itself accreted, $M_2'$ loses 90% of its stripped mantle remnants to $M_1$, explaining the missing mantle. The result is a dichotomy between 'accreted' and 'unaccreted' populations, the latter population highly diverse and including a dominant proportion of HRC survivors. These ideas are developed further below.

**Table 1.** Symbols and acronyms used in this review; see also **Figure 3**.
  **SSC** = similar size collision ($R_1 \sim R_2$, $v_{rel} \sim v_{esc}$),
  **HRC** = hit and run collision ($\xi \sim 0$)
  **NLB** = next-largest bodies, the most massive contributors to the largest bodies
  $N_{final}$ = last unaccreted NLBs, of $N \gg N_{final}$
  **GMC** = graze and merge collision ($\xi \sim 1$)
  **SFD** = size frequency distribution, $dn \sim R^{-\alpha} dR$
  $\xi$ = accretion efficiency, $(M_F - M_1)/M_2 \leq 1$
  $h$ = number of HRCs experienced by a body
  $a$ = accretionary attrition, $\ln(N/N_{final})$
  $\phi$ = scaled relative velocity $v_{rel}/v_{esc}$
  $\gamma$ = normalized projectile mass $M_2/(M_1+M_2)$
  $\theta$ = impact angle at contact, $\sin^{-1}[b/(R_1+R_2)]$
  $M_2'$ = identifiable remnant of $M_2$ (e.g. core)

**Final accretion**

In a typical *N*-body simulation of planet formation, a few dozen embryos (point masses *M* of radius *R*) orbit the Sun chaotically and collide. When two bodies' center-of-mass separation $r < R_1 + R_2$ they collide and can potentially merge. Collisional mergers in *N*-body simulations are found to produce Venus- and Earth-like planets, plus a few unaccreted objects representing Mercury and Mars. This 'late stage' of giant impacts (Wetherill 1985) can be extended to considering the final accretion of massive planetesimals such as Vesta and Ceres. Like the Earth, they might have accreted in tens or hundreds of massive but scaled-down SSCs.

Final accretion is often approximated as perfect merger, in simulations and theory. But the process is only half-efficient. For one thing, a gravitating body cannot acquire arbitrary angular momentum. $M_1$ is unable to hang on to all of the colliding materials, or equivalently, $M_2$ is not slowed down enough to be captured. Agnor et al. (1999) studied the assumption of perfect merger in *N*-body simulations, and found that it resulted in terrestrial planets spinning faster than $P_{rot} \sim 1$ hr, greatly exceeding the spin-disruption threshold (Chandrasekhar 1969). This is sometimes interpreted to suggest the viability of Moon-formation scenarios like Darwin (1876) starting with a proto-Earth spinning near the brink of disruption. But the real implication is that perfect mergers are unphysical.

The complexity of SSCs requires three-dimensional computational modeling. The most common approach is smooth particle hydrodynamics (SPH), originally applied to studies of the Moon-forming giant impact (e.g. Benz et al. 1989, Reufer et al. 2012). Early studies (Agnor and Asphaug 2004, Asphaug et al. 2006) modeled terrestrial planetary embryos colliding over a range of expected velocities and impact angles. They found that the limits on angular momentum acquisition place strong limits on mass acquisition. Rotation periods faster than $P_{rot}$~3-5 hours are difficult to attain, and about half of SSCs are found to be mantle-stripping or disruptive HRCs (see more recently Bonsor et al. 2015). Agnor and Asphaug (2004) suggested that halving the accretion inefficiency would double the timescale of planet formation. The first *N*-body study to include and track HRC remnants (Chambers 2013) found that the timescale extends to 160 Ma or longer, consistent with modern ideas for a late-forming Moon and a long tail of major collisions.

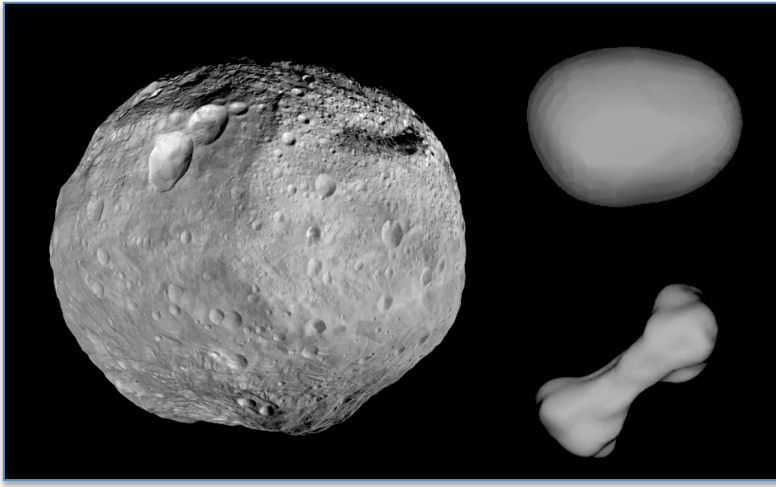

**Figure 1.** Clockwise from left, approximately to scale: (a) 4 Vesta, a 525 km diameter silicate asteroid with an iron core (NASA/Dawn). (b) 16 Psyche, the largest metallic asteroid, 240 x 185 x 145 km diameter, perhaps the disrupted remnant of a Vesta-like body (photometric model; Kaasalainen et al. 2002). (c) 216 Kleopatra is a 220 x 95 km metallic 'dog bone' (radar image; Ostro et al. 2000). According to the HRC hypothesis, Vesta may be an oligarch, whereas Psyche and Kleopatra are NLBs among the $N_{final}$.

Collisional accretion at the planetesimal scale is revealed in the largest asteroids of the Main Belt, whose diversity is even more extreme than the terrestrial planets. A principal distinction is that accretion in the Main Belt was lossy; extrapolating the protoplanetary disk between Mars and Jupiter suggests an original mass ~0.1-1 $M_\oplus$ (Weidenschilling 1977, Farinella et al. 1982). Of the 0.1–1% that survived, half the mass is in five 300–1000 km planetesimals that range in bulk density from ρ~2–4 g/cm$^3$. These represent 3 or 4 unique spectroscopic classes (e.g. Demeo et al. 2009) – a diversity that is completely unexpected for a narrow region of the nebula. The other half of the mass is a grab-bag of thousands of objects with myriad taxonomies, the more massive ones being probably primordial (Bottke et al. 2005) and the smaller ones representing an evolving collision cascade.

Collisional grinding cannot explain Main Belt attrition (Farinella et al. 1982), leaving two categories of ideas: (1) Most of the planetesimals accreted into Moon-to-Mars-sized bodies (as they did elsewhere) that consumed nearly all the mass. These planets grew fast enough to be ejected by a mass-dependent process (Chambers and Wetherill 1998; Ogihara et al. 2015), and the Main Belt was stranded. A thousand ~500–1000 km diameter original planetesimals might have accreted into one lost planet, according to this scenario, leaving Vesta and Ceres. (2) Planetesimals were scattered in a size-independent manner by giant planet migration (e.g. Walsh et al. 2011), quenching their accretion into planets (Bottke et al. 2005). Here Vesta and Ceres would be the last two of a few hundred oligarchs that were ejected by mass-independent resonant scattering, a random sampling of objects stalled the beginning of late stage accretion.

Each scenario would leave a distinct imprint, due to its specific manner of attrition. In the first scenario Vesta and Ceres are lucky NLBs that avoided being accreted by the long-lost planet before it got ejected. They are remnants of its major feedstock. However, by the time almost every NLB has been accreted by the lost planet, every unaccreted NLB has had on average a few HRCs, according to the attrition arguments below (**Figure 9**). But Vesta and Ceres are thought to be intact planets (Russell et al. 2012), not disruption relics.

In the second scenario, Vesta and Ceres are lucky in a very different way. It's not that they avoided being accreted; they avoided the forest of resonances that ejected all of their sisters. In this case they are two of the most massive bodies (oligarchs) that accreted in the original Main Belt. This scenario has better potential for explaining the perplexing diversity of ~200 km asteroids (Psyche, Hygiea, Interamnia, etc.) as these would be the final NLBs that suffered hit and runs during the accretions that formed dozens or hundreds of Vesta- to Ceres-size bodies.

## 2. Catastrophic Disruption

Direct evidence for catastrophic disruption of planetary progenitors is found in suites of meteorites (McSween 1999; Keil et al. 1994) including thousands of irons that are thought to sample exhumed cores of ~50–100 differentiated planetesimals (Wood 1964, Wasson 1990). Astronomical evidence for disrupted minor planets is less straightforward to interpret, because spectroscopy detects only surface characteristics, and asteroid densities are seldom measured. But at least a few major asteroids are thought to be metallic cores, including 16 Psyche, a ~200 km diameter spheroid, and 216 Kleopatra, a 90×220 km 'dog-bone' shaped object, and probably others (compared in **Figure 1**).

Given the apparent frequency with which cores have been exposed and exhumed, the impact disruption of massive planetesimals needs to be effective. But the fact is, the impact kinetic energy per unit mass $Q^*_D$ that is required to destroy and disperse a planet (so that final mass $M_F \leq \frac{1}{2}M_1$) increases disproportionately with target radius (Benz and Asphaug 1999; Leinhardt and Stewart 2012). Planetesimals larger than ~100–200 km diameter are effectively immune to impact disruption, for realistic encounter velocities. Evidence for this limit is indicated by the strong peak in the Main Belt differential SFD at ~100–150 km diameter according to O'Brien and Greenberg (2003) and Bottke et al. (2005).

**Stripping of mantles**

Extraordinary impacts are required to excavate core material from deep within large planetesimals, and that is the problem HRC was invented to address. According to the theory that $M_2$ beats $M_1$ to pieces, a basic requirement is to start with a Vesta-like $M_1$ and end with a Psyche-like remnant once it is destroyed. Most of the silicate mantle, ~3/4 the volume of $M_1$, must be ejected to escape velocity, and thereafter disposed of by a process that neither destroys the remnant core with an ongoing fusillade, nor causes the core to reaccrete most of its lost material.

Today there are ten >200 km diameter asteroids for every >500 km asteroid, so an impact between a 200 km projectile into a Vesta-size target is plausible. (This distribution could also represent how Vesta was accreted, as argued below, but at lower velocity.) For erosion scenarios we consider modern Main Belt impact velocities (5 km/s), or possibly twice that (10 km/s). Assuming a nominal impact angle (θ=45°) even the fastest of these does not eject core material (**Figure 2**). A head-on collision at 10 km/s, not shown, causes the escape of intensively shocked core material, but leaves behind tens of km of mantle silicates. Core exhumation is possible, but not probable, and not without intensively shocking the target asteroid.

The classic model requires extraordinarily energetic impacts because it requires shock acceleration to win out over gravitational acceleration. As in cratering, shock excavation is disproportionately difficult with size (the reason the largest impact basins are full of shock melted rocks; Melosh 1989). Shocks are observed in rocky materials for amplitudes ≳$10^{11}$ dyn/cm² (Rodionov et al. 1972), and shock amplitude decreases with the 2–3 power of distance in impacts, so shock acceleration is localized. To cause global disruption of a major planetesimal, shock acceleration must persist over ~100 km distances to reach the CMB, from where core material must push through an increasing amount of mantle against its gravitational potential. This inefficiency is represented by the steep slope of $Q^*_D(R)$ computed from simulations (Asphaug and Benz 1999; Leinhardt and Stewart 2012).

The requirement of mantle removal is made even more severe by the wide range in cooling rates recorded in iron meteorites, that require almost complete stripping of silicate mantle. When a metallic core solidifies inside an insulating silicate layer, it does so isothermally because of iron's high thermal conductivity, which is ~30 times greater than crystalline rock and >1000 times greater than lunar regolith. Yet metallographic cooling rates of IVAB iron meteorites span ~2 orders of magnitude, with slower cooling corresponding to higher nickel abundance and hence greater depth inside the core. Thermochemical modeling indicates a ~300±100 km diameter iron body (Yang et al. 2007) cooling beneath a ≲0.3 km silicate crust (Moskovitz et al. 2012). Not all iron parent bodies were stripped into bare metallic spheroids, but some apparently were.

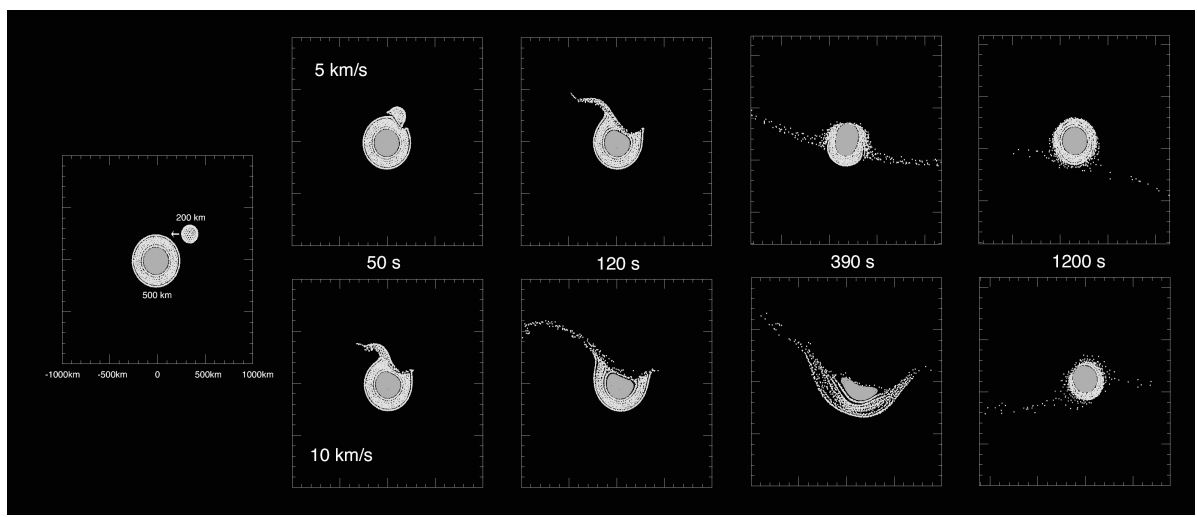

**Figure 2.** Attempts to exhume core material by asteroid collision. A Vesta-like asteroid $M_1$ (500 km diameter, 70% rock, 30% metal) is struck by a 200 km diameter projectile ($M_2$) that comes in from the right at 5 km/s (top) and 10 km/s (bottom), striking at θ=45°, the most likely impact angle. Light = basalt, dark = iron (planar slice of 3D SPH simulations; Asphaug 2010).

This raises a difficult point: Although HRC is much more effective than classic impacts at removing mantles, any single HRC still leaves behind an appreciable silicate layer. Even Mercury, with ~500 km of silicates outside a 2000 km iron core, is barely possible to achieve with a single HRC, and is better explained by a two-HRC scenario (Asphaug and Reufer 2014). So we have to account for multiple-HRCs for this approach to make sense.

**Disposal of rock**

One idea (Burbine et al. 1996) is that the Main Belt experienced wholesale mass loss after its formation by bombarding planetesimals. Silicate mantle, weaker than iron and less gravitationally bound, was more easily stripped away and beaten down to dust, goes the argument, at which point it was removed by solar radiation pressure and other effects. The idea has merit, although Farinella (1982) showed that grinding away ~99% lost material would damp the relatively fast rotations of the largest asteroids (Vesta and Ceres spin several times faster than any of the terrestrial planets). Also, because planetesimal disruption is a gravity-regime process (see below), an iron core exposed by this bombardment would be destroyed by the same cataclysm that just destroyed its mantle; the process would not stop with only ~¼ of the original mass remaining. And lastly, the 'battered to bits' scenario leaves us lacking an explanation for the preserved basaltic crust of Vesta, which should have been destroyed during this onslaught.

The problem of mantle stripping is compounded by having to explain where it all goes. Stony achondrite meteorites, that could represent mantle or crustal rocks, are much less common than irons and stony-irons among fallen meteorites. If Vesta-sized asteroids were catastrophically disrupted dozens of times, their remnants must have been ground down and transported away. Unless completely efficient, the removal process should leave behind easily identifiable materials in dynamically protected regions of the Main Belt, and there should be ample evidence for the transport of huge quantities of crystalline silicate dust. Whether solar radiation pressure and other effects could achieve this level of dust removal is another matter entirely.

In summary, a successful mantle-stripping model must (a) remove the mantles of dozens of major planetesimals, sometimes completely, (b) without invoking an energetic onslaught that would remove the crusts of Vesta and Ceres, and (c) explain how and where all this mantle rock went missing. In addition, (d) this core-exhumation must be possible under low-shock conditions. Only ~half of iron meteorites have been strongly shocked (Jaeger and Lipschutz 1967), and low shock mantle stripping may apply as well to Mercury, whose crustal volatile abundances are more Earth-like than Moon-like (Peplowski et al. 2011).

**Final bodies and attrition**

If $N$ is the number of similar-sized bodies (NLBs) in a starting population, not counting any that are ultimately scattered, and $N_{final} \ll N$ is the number that remain unaccreted by the largest bodies, then

(1)  $a = \ln(N/N_{final})$

is the magnitude of the attrition. Original bodies that did *not* get accreted either never collided with an oligarch, or else every collision was non-accretionary (HRC). This biases the survivor population to becoming dominated by HRC remnants.

Under simplifying assumptions (below), the average number $h$ of HRCs experienced by a typical unaccreted NLB is $\langle h \rangle \sim a$ (**Figure 9**). If 20 embryos accreted in the late stage, leaving two, then $a=\ln(20/2)\sim 2$, so one $h=2$–$3$ final NLB is expected, and one $h=0$–$1$ NLB. This is why Mercury and Mars are different according to Asphaug and Reufer (2014). If a few hundred ~500-1000 km diameter planetesimals accreted into some now-lost planet(s), leaving two, then Vesta and Ceres would themselves be the NLBs. According to the attrition they would have to suffer several HRCs ($a\sim 4$) on average. But they seem like intact protoplanets. The more consistent scenario is that Vesta and Ceres are not NLBs, but are largest bodies that (along with ~100 sister oligarchs) swept up most of their own NLBs, leaving behind an unaccreted collection of hit and run remnants.

## 3. Surviving Projectiles

Perfect merger is defined as $M_F=M_1+M_2$ where $M_F$ is the mass of the largest remnant. Accretion efficiency is the mass-fraction added ($0<\xi\leq 1$) or eroded ($\xi<0$) from $M_1$,

(2) $\quad \xi = (M_F - M_1)/M_2$.

HRCs are SSCs with $\xi \sim 0$, although as noted the outcomes are highly varied. The consequences to $M_2$ scale about inversely (Asphaug et al. 2006) with the normalized projectile mass

(3) $\quad \gamma = M_2/(M_1+M_2)$.

For common parameters, $M_1$ is massive enough to cause the disruption of $M_2$, yet unable to accrete all of its angular momentum. Because a factor ~2 in radius means a tenfold mass asymmetry, combined geometrical and gravitational effects make mergers difficult, and grazing events common. In impact cratering the contact plane equals the target, and gravity is a constant vector pointed down, so grazing is exotic. SSCs have a finite, overlapping geometry (**Figure 3**) which means that the outcome is dominated by gravitational and inertial effects instead of shocks; so grazing is the norm. The HRC scenarios for Moon formation by Reufer et al. (2012) and for Mercury formation by Asphaug and Reufer (2014) both have $\theta\sim 30°$–$35°$. These are steeper than average ($\theta_{ave}\sim 45°$) yet grazing in downrange motion. Moderately-oblique SSCs ($\theta\sim 45°$–$75°$) are grazing to the point of producing pinwheel spirals, and even in a $v_{rel}=0$ merger a percentage of this spiral arm material escapes (Asphaug and Reufer 2013) as novel bodies.

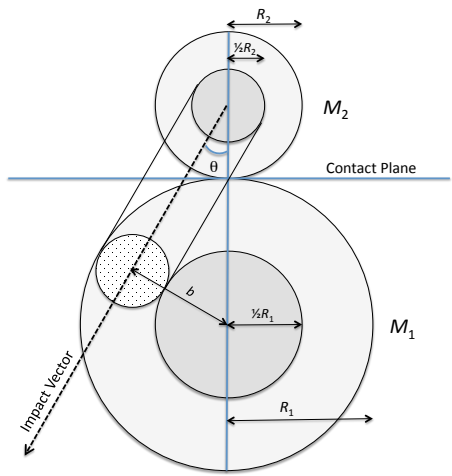

**Figure 3**. A typical SSC with $R_{core}=\tfrac{1}{2}R$. The cores do not intersect for $\theta>\sin^{-1}(\tfrac{1}{2})=30°$, and most of $M_2$ overlaps $M_1$. Because $\theta=45°$ on average, SSCs are usually grazing, hence the prevalence of HRC and GMC. There is no impact locus, and angular momentum dominates, so outcomes are evaluated using 3D simulations.

If most of $M_2$' is slowed and captured gravitationally, it is a graze and merge collision (GMC) as in scenarios of Moon formation (Benz et al. 1989; Canup and Asphaug 2001) and icy satellite formation (Canup 2008; Leinhardt et al. 2010; Asphaug and Reufer 2013). If identifiable remnants of $M_2$ escape, it is an HRC, which can be subdivided further based on whether $M_2$' is a core fragment, a shock vapor plume, a chain of clumps, and so on.

The transition between GMC ($M_1$ grows, $M_2$ vanishes) and HRC ($M_1$ is not much changed, $M_2$ is destroyed or dismantled) is a sensitive function of impact angle $\theta$ and the normalized pre-encounter relative velocity

$$(4) \quad \phi = v_{rel}/v_{esc},$$

where $v_{rel}$ is the relative velocity at 'infinity', and

$$(5) \quad v_{esc} = \sqrt{2G(M_1+M_2)/(R_1+R_2)}$$

is the two-body escape velocity, and

$$(6) \quad v_{imp} = \sqrt{v_{rel}^2 + v_{esc}^2}$$

is the impact velocity. In orbiting populations, instead of $v_{rel}$ we consider the random velocity $v_{rand}$ after subtracting the circular Keplerian velocity $v_{kep}$.

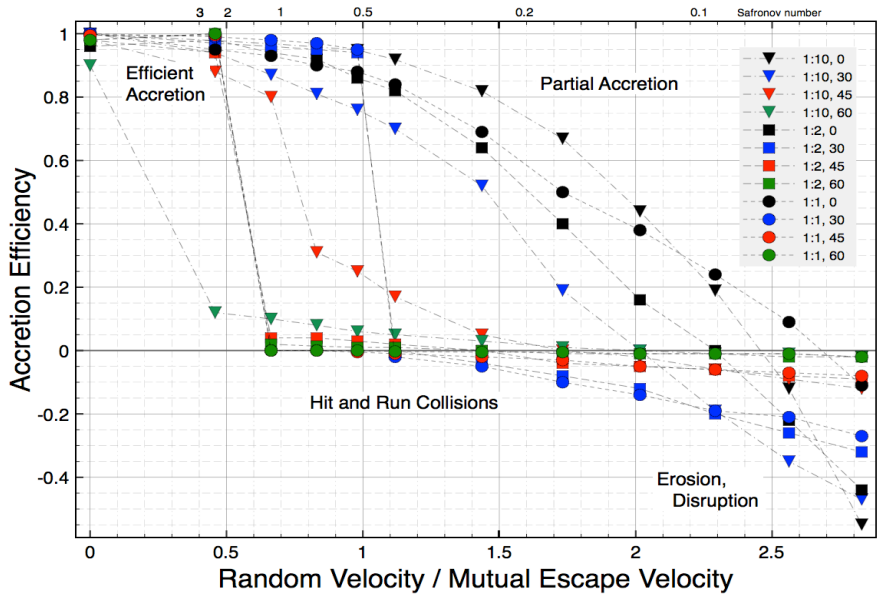

**Figure 4.** Accretion efficiency, ignoring returns, plotted as a function of normalized relative velocity $\phi$ (Asphaug 2010) and mass ratio $M_2:M_1$ as labeled. The target $M_1=0.1\ M_\oplus$ in each case. Self-stirred populations ($\phi\sim 1$) transition from accretion to HRC for impact angles $\theta\sim 30°$–$45°$ (colors). Because $\theta_{ave}\sim 45°$ either outcome is about equally likely.

Summary outcomes of terrestrial planet-forming collisions involving Mars-size targets are shown in **Figure 4**. The symbols are for mass ratios indicated by symbol shape ($\gamma$=1/11, 1/3, 1/2) and for impact angles $\theta$=0°, 30°, 45°, 60° indicated by colors. Accretion efficiency for single encounters (ignoring return events) is plotted over the velocity range 0<$\phi$<3, that is, nominal accretion velocities. The graph splits, with high-angle ($\theta \gtrsim$30°–45°) being HRCs for the most part, and low-angle ($\theta \lesssim$30°–45°) being accretion for the most part. Because the impact angle is random for a self-stirred population, the analogy of a coin flip can apply (see below).

Stripping or shredding of $M_2$ occurs in a process that scales effectively to massive planetesimals, and that causes catastrophic disruption during accretion, not long after. However, for planets or satellites orbiting a central body, HRC is the beginning of an extended dynamical and collisional interaction that is likely to end in further accretion – a drawn-out GMC. But likely does not mean always, and a fraction of the stripped or disrupted HRC material goes unaccreted. This leads to expanded diversity of remnants, and to complex genetic relationships, because orphaned asteroids can be stranded while their parent body is accreted.

## 4. Planetesimals to Embryos

In the absence of nebular gas, a self-stirred population (Safronov and Zvjagina 1969) becomes excited to $v_{rand} \sim v_{esc}$ due to close mutual encounters. In scenarios of oligarchic and late-stage growth (Kokubo and Ida 1998, Wetherill 1985), matter accretes by collisions involving bodies around the top of the size distribution, as represented by *N*-body simulations (e.g. O'Brien et al. 2006). We define SSC as the regime where objects comparable in size are accreting, that is,

(7) $\phi \sim 1$

(8) $R_1 \gtrsim R_2$

This includes giant impacts at tens of km/s, and 'small giant impacts' where planetesimals accrete at hundreds of m/s.

In this regime, collisional kinetic energy per unit mass goes as

(9) $Q \sim \gamma(\phi^2+1)v_{esc}^2$

where $v_{esc}^2$ is proportional to the gravitational binding energy G$M/R$ of the largest bodies. That is, the largest massive bodies can only be disrupted under high-$\phi$ conditions; then they grind down (Dohnanyi 1969) into small bodies which damp their motions. Feedbacks, especially around the strength-gravity transition where d$Q^*_D$/d$R$ changes from negative to positive (O'Brien and Greenberg 2003), produce bumps and wiggles in the SFD (Durda et al. 1998; Cuzzi et al. 2010).

Because there is no 'point source' to an SSC, crater scaling laws do not apply (Holsapple and Housen 1986). But they do obey hydrodynamic similarity. Departures from similarity arise when strength and friction and dilatancy are considered at small scales (Jutzi et al.

2015), and viscosity and compressibility, and then shocks at large scales. Phase transformation, mixing and reactivity can occur across all scales. These details are poorly understood, in part because the modeling of SSCs at ~10-1000 km diameter faces major computational challenges.

One challenge is to properly implement and validate the complexities of porosity, compaction, friction and cohesion into a numerical model (Jutzi et al. 2015; Asphaug et al. 2015). Another is that the computer timestep d$t$ decreases with spatial resolution, yet the simulation time is always a few times the gravity timescale

(10)   $\tau_G \sim (3\pi/G\rho)^{1/2}$,

several hours. So for small $R$ the number of timesteps becomes prohibitive. Another challenge is that a computational round-off error in density can translate into a pressure error exceeding the hydrostatic pressure, for small planetesimals with stiff EOS. Such errors can propagate and unphysical structures can emerge in simulations.

A kilometer-scale SSC is shown in **Figure 5**, where weak, porous cometesimals graze and merge at 40 cm/s (Jutzi and Asphaug 2015). Shown is for 0.5 kPa crushing strength, zero cohesion, dry friction, and impact parameters $\phi=1.1$, $\gamma=0.33$, $\theta=52°$. A benchmark for such studies is the disruption of comet Shoemaker-Levy/9, which suffered a tidal encounter with Jupiter, forming ~20 discrete sub-nuclei (Melosh and Scotti 1993; Asphaug and Benz 1994; Schenk et al. 1996). Accretions of small cometesimals result in layered piles (e.g. Belton et al. 2007). For comet-scale SSCs the boundary between GMC and HRC is found (Jutzi and Asphaug 2015) to depend on the impact angular momentum $L = \gamma v_{imp}(R_1 + R_2)\cos\theta$ divided by reference angular momentum $L_{\text{ref}} = \gamma v_{\text{esc}}(R_1 + R_2)\cos(45°)$, or

(11)   $L/L_{ref} = [2(\phi^2+1)]^{1/2}\cos\theta$

where HRC occurs for $L/L_{ref} \gtrsim 1.4$.

As planetesimals increase in mass, their accretions become more violent. Escape velocity in m/s equals $R$ in km (exactly for $\rho=1.9$ g/cm$^3$), so that cometesimals accrete at a walking-pace (**Figure 5**), and larger planetesimals accrete like a train wreck. For SSCs up to around 1000 km diameter, the impact kinetic energy is too small to involve shocks, but substantial frictional heating and crushing are expected at a global scale. During the first few Ma, $^{26}$Al heating might overwhelm other thermal effects. But accretion beyond a few 1000 km diameter involves shock-producing collisions, with the threshold shock stress depending sensitively on material composition and porosity. Jetting occurs in Moon-sized and larger collisions (Johnson et al. 2015). Bodies ~10–1000 km diameter are expected to acquire a strong petrographic imprint recording their history of collisions, depending on composition, porosity and thermal state.

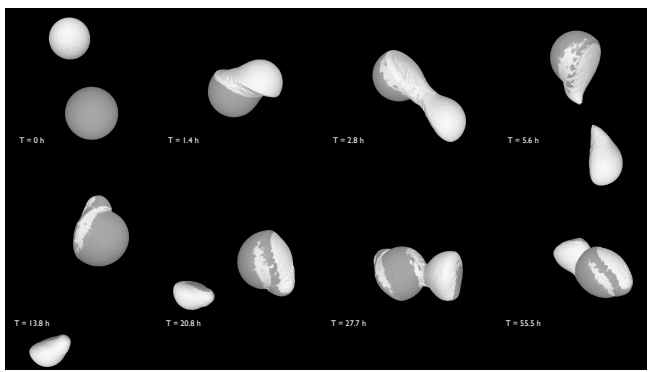

**Figure 5**. Graze and merge collision between km-sized porous ice spheres leaves a bi-lobed final body, in an SPH simulation with friction by Jutzi and Asphaug (2015). The bodies exchange material and re-collide ~24 hr later. Higher angular momentum collisions are HRC. Smaller projectiles can form layered structures instead of lobes.

There have been few spacecraft investigations of bodies ~30–300 km diameter, and all have been fast flybys. The best examples (see chapter by Russell et al.) are Vesta ($R$=260 km) and Ceres ($R$=470 km). Vesta is thought to have been globally melted, to produce its basaltic crust, but has substantial units of carbonaceous material (Reddy et al. 2012). As for next-largest asteroids, only 21 Lutetia (~90 km) and 253 Mathilde (~50 km) have been seen up close (Sierks et al. 2011, Weiss et al. 2012, Veverka et al. 1997). Lutetia is 2-3 times as dense as Mathilde and 5 times as reflective (Binzel et al. 1996, Barucci et al. 2012). Every asteroid in this size range seems special, and visiting a statistically meaningful sampling of major asteroids is an important science objective.

**Differentiation and segregation**

In our solar system, radioactive meltdown from $^{26}$Al→$^{26}$Mg decay ($\tau_{½}$=0.72 Ma) coincided with the planetesimal accretion timescale. This might not be the case in all solar systems. For estimated $^{26}$Al abundances, a 50 km planetesimal would have melted globally if it accreted in 1 Ma, and a 20 km diameter planetesimal would have melted internally if it formed in the first few 100,000 yr (e.g. Sahijpal et al. 2007). Heat production is balanced by convective cooling from the deep interior, limited by conductive cooling in case of a stagnant lid or crust. A melted planet insulated by a regolith could be encased in ice. The pace of cooling is recorded by the solidification sequence (Williams 2009; Bryson et al. 2015; Fu and Elkins-Tanton 2015).

Ongoing HRCs would strip the crusts and hydrospheres and atmospheres from unaccreted bodies, often repeatedly. Thus, in addition to producing a diversity of novel bodies, HRC evolution would greatly accelerate cooling whenever melted planetesimals are broken up and stripped of their cold exteriors (c.f. Ciesla et al. 2013). Being a gravity process, HRC removes a thick lid of crystallized materials as easily as it does a melted layer.

Besides causing accelerated cooling, stripping unloads the hydrostatic pressure deep inside. Consider a melted planetesimals 50 km diameter, with interior pressure $P_o$~10 bar. If released instantaneously this would be like an exploding car tire, but it is regulated by the inertia of the materials above. The thermal state of deep planetesimal magma is uncertain (e.g. Williams 2009; Fu and Elkins-Tanton 2015), and laboratory data for primitive silicate melts are only indirectly applicable, given that the acceleration of gravity is 1/1000 that of Earth and the timescales in question range from hours to >1 Ma.

Lack of knowledge leaves open some basic questions, such as core/mantle differentiation. As gravity increases, iron blobs can percolate through a solid matrix or drain through a magma ocean (Solomatov and Stevenson 1993). Percolation requires $\Delta\rho gr$ across a blob of radius $r$ to exceed the surface tension $s/r$. Assuming $s$~400 dyn/cm, and $\Delta\rho$~4 g/cm$^3$, and $r$=1 mm blebs, and $g$~1 cm/s$^2$, then $E_o \equiv \Delta\rho gr^2/s$ ~10$^{-4}$. For percolation to occur ($E_o \gtrsim 1$) blebs would have to grow to $r$~10 cm by e.g. shock, friction or vibration.

Further heating can break down the silicate matrix, forming a mush or magma ocean at $T \sim 1300\text{-}2100$ K depending on depth, composition and water content. Inside a melted body, assuming iron-silicate interfacial stresses are overcome, iron droplets can rain out on a Stokes timescale $\sim R\eta/gr^2\Delta\rho$. Here $\eta$ is the matrix viscosity, which depends sensitively on crystal content, volatile content, and bubble nucleation. These depend on $T$, $P$ of the evolving interior. For $\eta=10^4$ Poise, rain-out of 1 mm droplets happens in $10^4$ yr, while for $10^{-2}$ Poise it happens in days. A crystal-rich or bubble-rich magma ($\gg 10^{10}$ Poise) remains colloidal, while a degassed, intensively heated magma might drain, completely or partially.

Cores accrete more specific angular momentum than the silicate mantles they displace, inducing a differential rotation $\delta\omega_{rot} \sim 0.1$ hr$^{-1}$ across the core-mantle boundary (CMB). The corresponding shear velocity $v_s$ might overcome gravitational stratification and disrupt the boundary layer for wavenumbers $kv_s^2 > g(\rho_2^2-\rho_1^2)/\rho_1\rho_2$ or $k \sim 2g/v_s^2$, the Kelvin-Helmholtz limit for inviscid fluids. The CMB is predicted to be unstable at wavelengths $\sim R/10$ for this $\delta\omega_{rot}$. Other factors, such as mechanical stirring during subsequent accretion (Golabek et al. 2014), might further hinder effective differentiation as planetesimals grow.

Evidence for varying degrees of differentiation is found in the great compositional diversity of chondrites (Scott and Krot 2003). Stony-iron meteorites, also relatively common, are thought to represent kinetically-mixed CMB materials inside melted planetesimals. Mesosiderites are proposed by Haack et al. (1996) to be liquid iron core materials mixed intimately with cold outer silicates. But an extreme impact is required to exhume the core (e.g. **Figure 2**) and the shock would melt the silicates. HRCs is a gentler mechanism for core-crustal mixing. Similarly, the juxtaposition of large crystalline olivine and metallic iron within Pallasites can be achieved by HRC without intensive shocks.

**Planetesimals, embryos and shocks**

A planetesimal can be defined as a terrestrial body $\sim 10\text{-}1000$ km in radius. Bodies smaller than this, unless melted, have strength and friction effects subject to very different physics (Jutzi et al. 2015). Bodies larger than 1000 km are characterized by the onset of shocks during their accretion, specifically,

(12)   $v_{imp} = (\phi^2+1)^{½} v_{esc} > c$

where $c$ is the sound speed, several km/s for relevant materials.

For shocks to occur, $v_{esc} \sim 1$ km/s, or $R \sim 1000$ km, whether for silicate or hydrous materials since generally sound speed increases with density. Accordingly one can define planetesimals as bodies that accreted without shocks, and embryos as bodies whose accretion required shocks, each associated with discernible epochs of petrologic evolution, recorded in the chronometry of shock resetting (e.g. Sharp and DeCarli 2006).

# 5. Simulations of Hit and Run

Computer simulations show that SSCs are sensitive to initial conditions, forming escaping tidal arms, fragment chains, stripped cores, and mantle sheets. The sensitivity adds a stochastic element to planet formation, especially the abrupt jump from GMC to HRC in

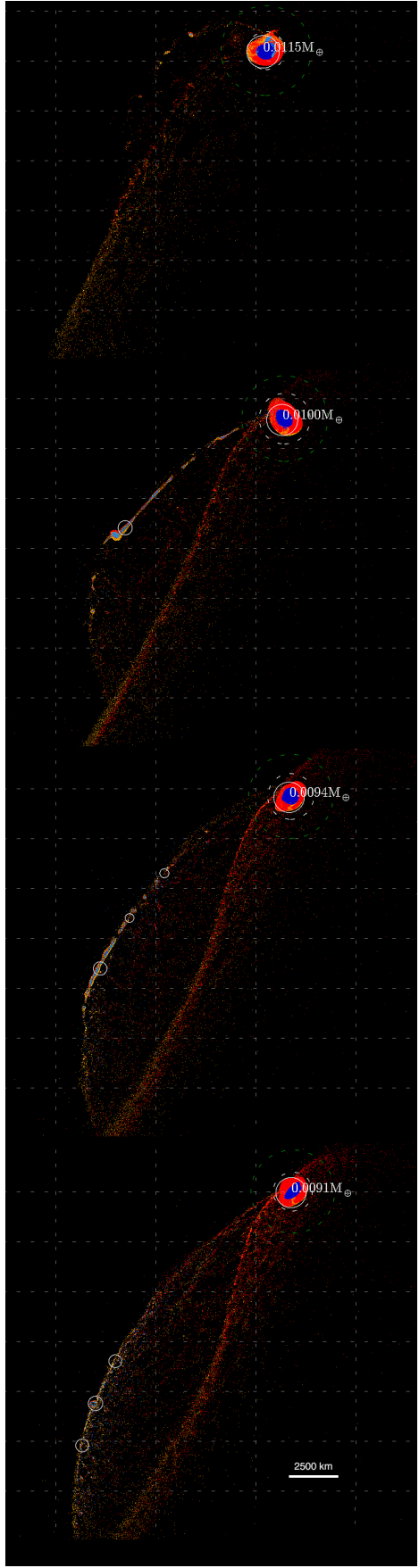

**Figure 6.** Collisions between large differentiated planetesimals, $M_1$=0.01 $M_⊕$, γ=1/6, θ=30°, $R_1$=700 km, $R_2$=400 km, at relative velocity ϕ=1.5, 2.0, 2.5, 3.0 (top to bottom, a-d). Note the changes in clump composition and morphology, and the segregation of material represented as Fe (blue) and $SiO_2$ (red). Shown are planar cross-sections; the iron-rich plume is an arm, while the silicate-rich "second wave" is a clumpy sheet. Solid circles represent gravitationally bound clumps (equivalent sphere). Green dotted circle plots the Roche limit around the finished mass, $M_F$ = 0.0115, 0.0100, 0.0094, 0.0091 $M_{Earth}$. White dotted circle is the corotation radius around $M_F$. (a) Partial accretion, ξ=0.75, shown ~4 hr after impact. (b-d) hit and run, ξ=0.00, -0.30, -0.45 respectively, ~2 hr after contact. Increasing spray from $M_2$ are brought into the mix whereas escaping core comes entirely from $M_2$. Shown are for θ=30°. Head-on collisions (θ~0) induce greater stress amplitudes while high-angle mergers have pronounced angular momentum effects, such as multiple spiral arms.

**Figure 4** – the difference between effective accretion, and the escape of a mantle-stripped planet or family of objects.

**Figure 6** shows a subset of calculations using the SPH code of Reufer et al. (2012) and the ANEOS equation of state for $SiO_2$ and Fe (Melosh 2007; Thompson and Lauson 1972). Two differentiated planetesimals ($R_1$=400 km, $R_2$=700 km, γ=1/6) collide at increasing speed, from 0.9 km/s to 2.8 km/s, θ=30° in each case. The physical resolution is ~50 km (smoothing length ~20 km, 200,000 particles). These collisions are subsonic. Generally, the mantle of $M_2$ goes one way (red clumps and sheets) and the core goes another (blue clumps), a recipe for segregating original planetary bodies into metal-worlds, ice-worlds, rock-worlds, and things inbetween.

## Segregation and clumping

GMCs and HRCs can produce discrete chains of middle-sized objects in clumped spiral arms, composed mostly of crustal or upper mantle material from $M_2$, plus material from deeper inside of $M_2$. Asphaug and Reufer (2013) model the middle-sized icy satellites of Saturn, ~300-1500 km diameter and compositionally diverse, as leftover clumps after

Titan accreted from an original satellite system that became unstable. Silicate-rich moons like Enceladus came from deeper inside of $M_2$ than ice-rich moons like Tethys; see also Sekine and Genda (2013). The Haumea system is proposed by Leinhardt et al. (2010) to be the result of a GMC associated with its satellites and the extended dynamical family.

**Transitions to giant impacts**

As noted, there is a gradual regime transition from planetesimal-scale to embryo-scale collisions, where shocks begin to dominate the physics and thermodynamics of the outcome. **Figure 7** shows a giant HRC that has identical parameters as **Figure 6d** but with $M_1=M_\oplus$ (ten times the size). Similarity breaks down because the giant impact at $v_{imp}$=30 km/s causes the effective transformation of kinetic energy through shocks into material entropy. Only one bound clump is identified (white circle), although this does not take into account gas physics that might limit dispersal.

The transition to giant impacts is recorded in meteorites. Krot et al. (2005) argue that chondrules in CB meteorites condensed from a shock vapor plume, but these are unusual. More common chondrules, the ~mm-sized porphyritic spherules found in most chondrite meteorites, crystallized under ~hours-long cooling times and retained semi-volatiles, inconsistent with condensation from expanding shock vapor. The prevailing model is that of shock melting of fluffy parcels in the dusty nebula, e.g. Morris and Desch (2010).

Johnson et al. (2015) propose an embryo accretion scenario for chondrule production, simulating interfacial jetting at the contact zone, assuming 2D symmetry is applicable. The embryos in this model must be primitive and highly porous in order to form massive jets of suitable composition, and have to be almost Moon-size in order for $v_{esc}$ to attain the required collisional velocities (~2 km/s). It remains to be understood why primitive porous embryos would exist, where they disappeared to, and how they avoided complete meltdown according to the thermal models.

**Pressure unloading**

Gravity creates a hydrostatic pressure that is released during HRC. The characteristic hydrostatic pressure inside a spherical planetesimal of density ρ is

(13)    $P_o \sim G\rho^2 R^2$

so $P_o$~10 bar inside a 50 km diameter planetesimal, and ~1-10 kbar inside the major asteroids. During inefficient accretion the original hydrostatic pressure is unloaded as unaccreted matter expands into space. In the limit of a barely-grazing SSC, pressure unloading is about 40% throughout $M_2$ (Asphaug et al. 2006), caused by tidal deformation alone. Apart from the mechanical and rheological stresses and shocks that can dominate the outcome, tidal unloading is substantially greater in an HRC because $M_2$ swings closer to the core of $M_1$. After that its materials disperse downrange to final pressure $P_f \ll P_o$, most of the mass evidently in one or a family of escaping clumps.

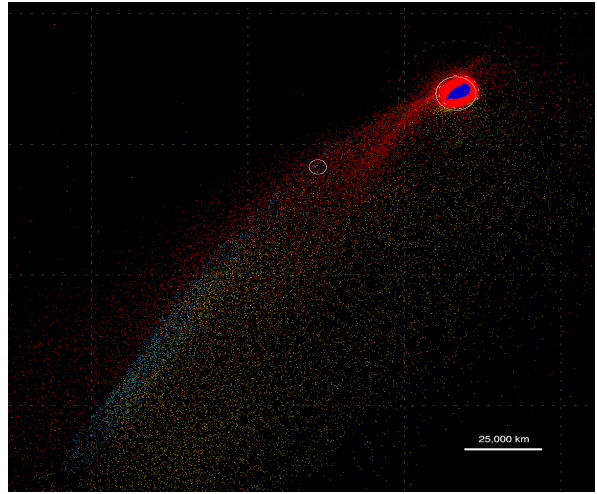

**Figure 7**. A giant impact HRC, identical to Figure 6d except the colliding bodies are 1000 times as massive, $M_1=M_\oplus$ (Asphaug and Reufer 2014), 2 hr after contact, $\gamma=1/6$, $\theta=30°$, $v_{imp}=3v_{esc}$. Instead of a chain of gravitating clumps, this supersonic collision features intense global shocks and expansive pressure unloading. White circle is the effective diameter of an identified gravitational clump.

25,000 km

One venerable idea (Sorby 1864) is that chondrules are produced in a process analogous to a terrestrial volcanic eruptions (Alidibirov and Dingwell 1996) or by impact splashing (Sanders and Scott 2012). However, planetesimal volcanism is a regime unfamiliar to experiments, involving a relatively unknown primitive material, expanding into near-vacuum conditions (Wilson and Keil 1982) at decompression rates of hours. And instead of being driven by gas expansion, the driver is the incoming velocity of $M_2$.

Droplets encapsulate an internal pressure $P_s$ due to their surface tension $s$, according to the Young-Laplace relation

(14)     $P_s=2s/r$

where $r$ is droplet radius. For immiscible fluids dispersing freely, $r\sim 2s/\Delta P$. Volcanic melts are not immiscible. Assuming some fraction $f$ of the unloading enthalpy goes into surface area production a.k.a. droplets; then

(15)     $r = 2s/fG\rho^2R^2$

is the droplet radius. Johnson et al. (2015) assert that decompression heating (gas expansion, exsolution, crystallization; Blundy et al. 2006) consumes the enthalpy and that nothing is left for surface area production. Reasoning that exsolution and gas expansion cannot happen without surface area production, Asphaug et al. (2011) argue that $f\sim\frac{1}{2}$. They show that $f=1\%$ is sufficient to produce millimeter-size droplets for ~100 km SSC. They focus on ~30 km planetesimals as these are large enough to be melted by $^{26}$Al yet might remain primitive when melted due to their micro-gravitational stress.

**Figure 8** shows a slice through a 30 km diameter partially differentiated planetesimal disrupting against a 70 km target, producing a spray of ~10 μm droplets according to Eq. 15. Other configurations yield similar results. The presence of a 5 km cold fragmented lid does not substantially change the dynamical outcome (Asphaug et al. 2011) but mixes crystalline silicates and possibly ices into the plume. Because the size $r$ of magmatic fragmentation scales with $\sim 1/R^2$, the question is not whether droplets can form from plumes unloading from the interiors of melted planetesimals – this is clear – but whether incompletely differentiated melted interiors can exist. If they cannot, then inefficient accretion should be a factory for basaltic spherules, which are not observed, instead of primitive chondrules which are ubiquitous.

## 6. Accretion and Attrition

Asteroids are 0.03% the mass of the terrestrial planetary zone. Mars and Mercury are 8%. These populations are residues, a last surviving feedstock subject to attrition and its biases. A simple model is developed, not a dynamical analysis but a statistical framework for why there is strong diversity in unaccreted populations.

About half of SSCs are accretionary (**Figure 4**); for simplicity call these mergers, ξ=1. The other half are non-accretionary, in some form or other; for simplicity call them bounces, ξ=0. Recalling **Figures 6-8** this is highly simplistic. Assuming this, and that collisions occur randomly (also highly simplistic), then accretion can be treated like a coin game where $N$ pennies are scattered on a table, with a few jars (representing potential oligarchs) placed among them. Select a penny at random and flip it; if it is heads, mark it $h$=1 (or HRC) and put it back on the table. Tails, put it in the nearest jar.

Continue flipping, and each time it is heads, mark it $h$+1, otherwise it disappears into the jar. After about $N\ln(N)$ random flips, most of the pennies are accumulated. The last pennies $N_{final} \ll N$ were either overlooked (never picked up, $h$=0) or else were flipped repeatedly and each time ($h$=1, 2, …) landed heads. Neither outcome is probable, seen individually, but the attrition bias is obvious: everything probable is in the jar. If a similar attrition bias applies to accreting planetesimals, it would bias the asteroids and meteorites to being remnants of repeated hit and run collisions.

Adopting this analogy, **Figure 9** shows the probability distribution function (PDF) for the $h$-number of an evolving population of $N$ bodies accreting by random collisions forming largest bodies. Each PDF ranges from the few bodies that avoided all collisions and remain primitive (*h*=0) to the half that have suffered more than $\langle h \rangle \sim \ln(N/N_{final})$ hit and runs in a row. The width of the distribution increases with attrition, which causes increasing diversity because of the wide ranges of collisional byproducts.

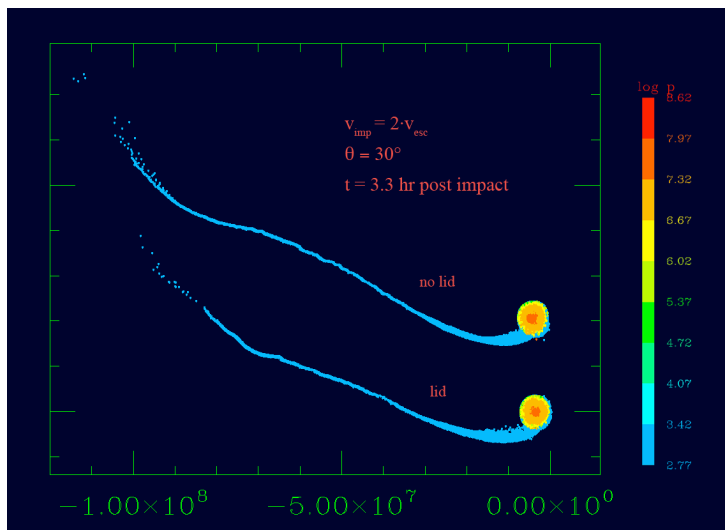

**Figure 8**. Proposed chondrule formation in an HRC between 30 km and 70 km diameter partially-differentiated planetesimals, $v_{imp}$=2$v_{esc}$=72 m/s, θ=30°, γ=0.07. Pressure in log dyn/cm$^2$ (~0.6 mbar to ~0.6 kbar). Little entropy is produced in the collision. Most of $M_2$ that becomes an expansive sheet (blue, millibars) producing droplets $r \ll 1$ mm according to Eq. 15. Top is for bodies without a cold lid; below is for a 5 km lid of the same material that includes dry friction (e.g. a crust). From Asphaug et al. (2011).

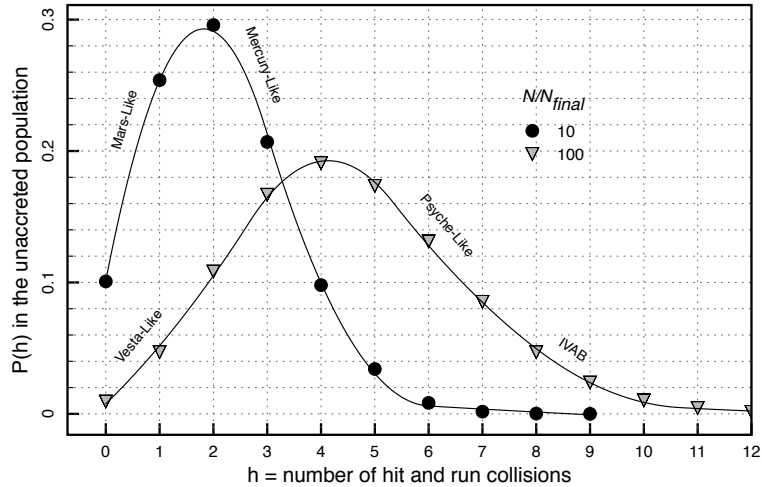

**Figure 9.** When *N* bodies accrete pairwise at random, they leave behind $N_{final}$ unaccreted bodies that have a probability *P(h)* where *h* is the number of HRCs experienced (averaged over thousands of trials). Maximum probability is for $h \sim a$, and the increasing spread in *h* shows that relatively original relics can exist alongside repeatedly-stripped remnants. Survivors with $h=0$ never interacted with a larger body. Repeated HRC could be common.

This was proposed by Asphaug and Reufer (2014) to explain why Mars and Mercury are so fundamentally distinct. If 20 terrestrial planets accreted, forming Earth and Venus, leaving Mercury and Mars as random leftovers, then $a \sim 2$. About half of the time a leftover would be "Mars-like" (*h*~0–1) and half the time it would be "Mercury-like" (*h*~2–3). Similarly, for asteroids, if Ceres accreted out of ~30 original ~300 km bodies, then $a \sim 3$ and outcomes could be "Vesta-like" (*h*~1) or "Psyche-like" (*h*>3).

The statistical representation compares reasonably with Chambers (2013), the only published *N*-body simulations to track HRC survivors explicitly. The simulations begin with *N*=14 embryos, each larger than Mars; idealized outcomes are applied for HRC and fragmentation, and the inner solar system is truncated at 0.2 AU, biasing against remnants near Mercury. Dynamical friction is contributed by 140 smaller bodies. Embryos evolve until there are (usually) two major planets, although taking longer than without HRC. Mercury- to Mars-size bodies are very underrepresented in the outcomes, because the starting embryos are larger than Mars, but nevertheless in 1 of 8 simulations, Chambers (2013) reports one stable embryo that suffered four HRCs, the last one similar to the scenario modeled by Reufer and Asphaug (2014) for a mantle-stripped Mercury. According to **Figure 9**, *h*=4 should be experienced by 10% of the $N_{final}$, which is not far off the mark.

**Hiding in Plain Sight**

Given that the mantles of Mercury and the metallic asteroids have disappeared, and that there are few rocky achondrites corresponding to the plethora of iron meteorites, there is a 'great dunite shortage'. This is explained by a return phase, as HRC remnants orbit the Sun or central planet, and accrete and reaccrete whatever debris remains within their sphere of influence (e.g. Jackson and Wyatt 2012). During this phase the orbits are intersecting, and the rate of sweep-up is a body's velocity through a swarm times the swarm density, times the geometric cross section $\pi R^2$, times the gravitational focusing factor

(16)  $f = 1 + \phi^{-2}$

Assuming uniform swarm density, and that $v_{rel}$ of the swarm is constant, then the sweep-up is biased to favor the largest, highest-$v_{esc}$ (largest $R$ and $f$) body. Computing the relative cross sections for $\gamma \sim 0.1$, the prediction is that ~90% of the materials stripped from $M_2$ accrete onto $M_1$. Mantle-stripped silicates ultimately end up on the target, augmenting its silicate composition by a few percent, which goes unnoticed. Seen as an aggregate process this is analogous to Ostwald ripening (Voorhees and Glicksman 1984) so that the unaccreted population systematically loses silicates and other exterior materials.

**Hit and run return**

This means that **Figure 4** is not a complete representation, because in most cases $M_2$' will itself be accreted by $M_F$, a hit and run return (HRR). This is analogous to GMC except that the return can take thousands or millions of years and appear like any other impact. In other words the same coin gets flipped again, and perhaps again (e.g. selections are not random). If two bodies are fated to accrete once they collide, no matter how many returns are required, then accretion is effectively perfect and the $N_{final}$ never collided with anything ($h=0$), contrary to the ideas presented above for disruption and mantle stripping and silicate disposal. Otherwise, a percentage of the interacting bodies must become isolated following a collision, the way Mercury must somehow (with ~1:10 probability) disentangle from proto-Venus or proto-Earth in the HRC scenario, in basic agreement with $N$-body simulations (Chambers 2013). In these cases the $h$ distribution might resemble **Figure 9**.

According to Reufer et al. (2012), the Moon may be a consequence of HRC, when Theia ($M_2$) collided into proto-Earth. They also consider Mercury-like and Ganymede-like projectiles, among other possibilities. Colliding closer to head-on than the standard model ($\theta \sim 30°$) and faster ($v_{imp} \sim 1.25\, v_{esc}$), the HRC contributes more entropy, heating the outer mantle and silicate atmosphere to 10,000 K. But most of Theia is left orbiting the Sun in this scenario, so the corollary is a probable second giant impact into Earth, or impacts by disrupted clumps. Return giant impacts, thousands of orbits later, strike at random $\theta$ and could offset the Earth-Moon inclination. Moreover, return giant impacts could augment a smaller moon formed in the first HRC, piling on Earth-derived silicates. And not all HRCs end in return: one can entertain the scenario that Mercury *is* Theia, isolated dynamically after colliding with Earth and forming the Moon.

Looking towards specific multiple-HRC or HRR solutions for Mercury and the Moon is tempting, and feasible, recycling the output of one simulated collision to initiate another. But further research is stymied by the fundamental question of where to begin. Impact variables $\theta$, $\gamma$, $\phi$ and $v_{esc}/v_{kep}$ are compounded by the same choices for the second collision, implying thousands of distinct scenarios, even ignoring pre-impact rotations and compositions. It requires fine-tuning of wide-ranging scenarios. The study of planetesimals, with much larger $N$, is a better determined problem at present, explaining classes of bodies through statistical arguments, but here too there is much work to be done.

# 7. Conclusions

Hit and run was hiding inside the assumption of perfect merger, and its consequences were masquerading as ballistic disruption – a process that requires intense shocks even at modest scale, and appears impossible for planetesimals >100-200 km diameter. Planetesimals of any size are readily disrupted by HRC so long as there are larger bodies trying to accrete them. The statistical argument of attrition shows that HRCs may not only be common in the formation histories of planetesimals, but prevalent.

Venus and Earth contain 92% of the inner solar system's mass and are compositionally similar. The next-largest bodies, Mercury, the Moon and Mars, contain 8% and are famously diverse. Asteroids are even more diverse, and if primordial, require many distinct accretionary environments within a narrow region of the protoplanetary disk. HRC can create this compositional diversity when differentiated or partly differentiated bodies are dismantled by non-accretionary collisions: core goes one way, crust and mantle another, hydrosphere another. Some of these remnants attain dynamical separation, becoming surviving cores or orphans, retaining the isotopic signatures of $M_2$ but having diverse bulk compositions. Generally the stripped materials orbit the Sun and mostly reaccrete onto $M_1$, explaining, at least statistically, the missing mantle paradox and core-rich remnants.

Every unaccreted planetesimal is unaccreted in its own way, increasing the diversity of leftovers. Most of the $N_{final}$ in an accreting population end up surviving one or more HRCs, each outcome sensitive to the specific parameters of the collision. Some of the $N_{final}$ are lucky to have never encountered a larger body, and the model predicts that a fraction of them ($\sim N_{final}/N$) are relatively primitive. The typical unaccreted object evolves to $h \sim a = \ln(N/N_{final})$, consistent with the idea of multiple-HRC origins.

In addition to causing global disruption and segregation, HRC can cause the thermodynamic transformation of the bulk of a planetesimal, even at scales and velocities where shocks are unimportant. Materials deep inside of growing planetesimals are unloaded from ~10-100 bars of hydrostatic pressure into clumps, sheets and droplet-swarms, depending on pre-encounter composition and temperature, and impact parameters. Alteration and degassing are initiated as volatiles flow along new pressure gradients, as bodies are stripped abruptly of massive insulating crusts and are clumped into novel, smaller bodies. These possibilities, and the strong statistical bias towards HRC survivors, provide important context for the petrologic interpretation of meteorites.

Vesta and Ceres appear to have avoided repeated HRCs with larger embryos, having retained their outermost layers (e.g. Clenet et al. 2014). Their silicate- and ice-rich compositions are inconsistent with them being the final feedstock of a lost Main Belt planet, because the prediction for strong accretionary attrition ($a \sim 4$) would make them subject to repeated mantle stripping. A more consistent story is for Vesta and Ceres to be two of the largest planetesimals that accreted in the region, the last of ~100–1000 oligarchs that were mostly scattered. In that case, the remarkably diverse ~100–300 km asteroids and their meteorites are from the mantle-stripped interiors and orphaned mantles and crusts left over when a primary population of ~300–500 km bodies were mostly but not completely accreted to form the ~500–1000 km diameter largest Main Belt planetesimals.

# Acknowledgements

This effort was sponsored by NASA NNX13AR66G, "Collisional Accretion of Similar-Sized Bodies". I am especially grateful for the diligent efforts of two anonymous referees.